\documentclass[12pt]{iopart}

\usepackage{amsmath}
\usepackage{amsfonts}
\usepackage{amssymb}
\usepackage{bm}
\usepackage{siunitx}
\usepackage{tabularx}

\usepackage{graphicx}%
\usepackage{multirow}%
\usepackage{amsthm}%
\usepackage{mathrsfs}%
\usepackage[title]{appendix}%
\usepackage{xcolor}%
\usepackage{textcomp}%
\usepackage{manyfoot}%
\usepackage{booktabs}%
\usepackage{algorithm}%
\usepackage{algorithmicx}%
\usepackage{algpseudocode}%
\usepackage{listings}%
\usepackage{mathtools}
\usepackage{siunitx}
\usepackage{cancel}
\usepackage{ulem}
\usepackage{subcaption}
\usepackage[numbers]{natbib}
\usepackage{natbib}
\usepackage{hyperref}

\usepackage{import}
\usepackage{url}

\newcommand{\mymtx}[1]{\textcolor{black}{#1}}
\newcommand{\sw}[1]{\textcolor{black}{#1}}
\newcommand{\et}[1]{\textcolor{black}{#1}}

\theoremstyle{thmstyleone}%
\theoremstyle{thmstyletwo}%

\theoremstyle{thmstylethree}%

\begin{document}

\title[Supervised Graph Contrastive Learning for Particle Decay Reconstruction]{PASCL: Supervised Contrastive Learning with Perturbative Augmentation for Particle Decay Reconstruction}

\def \Ours {PASCL}
\def \Code {https://github.com/lukSYSU/PASCL}
\def \Data {https://doi.org/10.5281/zenodo.6983258}

\author{Junjian Lu$^{1,2}$\footnote{Both authors contributed equally to this work.}, Siwei Liu$^{2,\ddagger}$,
Dmitrii Kobylianski$^3$, Etienne Dreyer$^{3,\S}$, Eilam Gross$^3$, Shangsong Liang$^{1,2}$\footnote{Corresponding authors.}}

\address{$^1$ Sun Yat-sen University, China}
\address{$^2$ Mohamed bin Zayed University of Artificial Intelligence, UAE}
\address{$^3$ Weizmann Institute of Science, Israel}
\ead{lujj36@mail2.sysu.edu.cn, siwei.liu@mbzuai.ac.ae, dmitrii.kobylianskii@cern.ch, etienne.dreyer@weizmann.ac.il, eilam.gross@weizmann.ac.il,  liangshangsong@gmail.com}

\begin{abstract}
In high-energy physics, particles produced in collision events decay in \sw{a format of} a hierarchical tree structure, where only the final decay products can be observed using detectors. However, the large combinatorial space of possible tree structures makes it challenging to recover the actual decay process given a set of final particles. \sw{To better analyse the hierarchical tree structure, we propose a graph-based deep learning model to infer the tree structure to reconstruct collision events.}
\sw{In particular}, we use \sw{a compact matrix representation termed as \textit{lowest common ancestor generations (LCAG) matrix},  to encode the particle decay tree structure}. Then, we introduce a perturbative augmentation technique applied to node features, aiming to mimic experimental uncertainties and increase data diversity. We further propose a supervised graph contrastive learning algorithm to utilize the information of inter-particle relations from multiple decay processes.
Extensive experiments show that our proposed supervised graph contrastive learning with perturbative augmentation (\Ours{}) method outperforms state-of-the-art baseline models on an existing physics-based dataset, significantly improving the reconstruction accuracy. \mymtx{This method provides a more effective training strategy for models with the same parameters and makes way for more accurate and efficient high-energy particle physics data analysis.}
\end{abstract}

\noindent{\it Keywords}: graph neural network, supervised contrastive learning, data augmentation, particle reconstruction


\maketitle



\section{Introduction}\label{sec:introduction}
\begin{figure}[t]
    \centering
    \includegraphics[width=\linewidth]{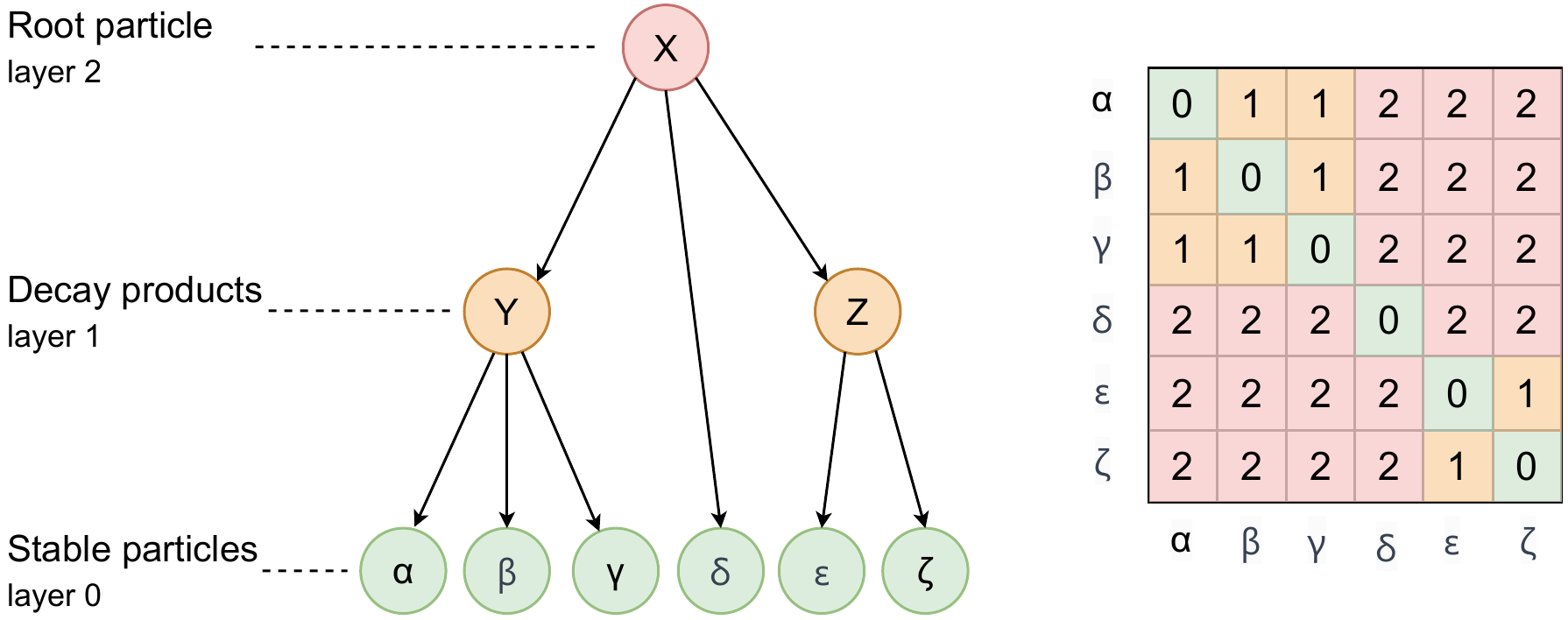}
    \caption{A simulated particle decay event from the Phasespace dataset with 6 stable particles (a modified version of figure 2 in~\cite{LCAG}). The tree structure shows an example of particle decay process (left), and the matrix represents its corresponding LCAG matrix (right). Labels and colors are only used as a symbol for discrimination.}
    \label{task_study}
\end{figure}

\et{Collider experiments such as the detectors at the Large Hadron Collider \cite{LHC} and Belle II at the SuperKEKB accelerator \cite{Belle-II:2010dht} shed light on the physics of fundamental interactions by recording data from particle collisions. Before reaching the detector, most of the particles produced in a collision decay into new, intermediate particles. Some of these are stable, while others likewise decay, and so on. The entire process can be described as a tree branching outward until the final set of stable particles, represented by leaves. An example is shown in \fref{task_study} (left). Reconstructing these particle decay trees essentially works backward to the collision itself and helps determine the physical process that generated the particles. A prime example is processes initiated by bottom and top quarks, which give rise to characteristic decay topologies caused by the unique properties of the intermediate particle states. However, more common processes such as signatures produced by light quarks and gluons can also be probed from the point of view of a decay tree. The ability to reconstruct decay trees can benefit jet classification \cite{Topograph} overall event reconstruction \cite{set2tree_ref7}.}


\sw{In recent years, machine learning (ML) has become increasingly influential in particle physics, as demonstrated by various studies~\cite{jet_rec_ref1,jet_rec_ref3,hepmllivingreview}. Graph neural networks (GNNs), in particular, have been especially suitable to a range of tasks in this domain, as reviewed in~\cite{jet_rec_ref4,DeZoort:2023vrm}.}
Given their ability to handle set-valued data and capture spatial relationships, GNNs have been highly applicable to particle tracking~\cite{jet_rec_ref10,jet_rec_ref11} and reconstruction~\cite{Qasim:2021hex,Mokhtar:2023fzl,jet_rec_ref13}.
Mostly, these tasks have focused on predicting the set of final, stable particles that reach the detector, \sw{but} not the full decay tree preceding them.
Likewise, GNNs have proven successful in tasks probing the decay tree inside of jets, such as finding secondary vertices~\cite{set2tree_ref2} and detecting substructure~\cite{set2tree_ref1,set2tree_ref7}.
These studies have been limited to specific topologies through assumptions of the underlying process (e.g. bottom-quark jets and boson jets) and are not designed to generalize to a broader range of possible decay trees.

\et{Recently, the authors of~\cite{LCAG} have introduced a new, compact, hierarchical tree representation known as the \textit{lowest common ancestor generations} (LCAG) matrix that serves as a well-defined target that does not assume the tree structure \textit{a priori}, avoiding the above limitations. In~\cite{set2tree_ref8}, a related approach was used in a more specific physics application, namely heavy-flavor hadron decays. Both papers likewise report some tendency for overtraining and acknowledge the possibility of future improvements to the trained model. We therefore aim to build on these existing efforts by incorporating techniques promoting model robustness and performance generalization, specifically, supervised contrastive learning with data augmentation.}

Data augmentation is a technique in machine learning that is used to reduce model overfitting by training a model on several copies of existing data that have been slightly modified. As a result, data augmentation effectively enlarges the training set without imposing significant computational costs. However, how to effectively generalize the concept of data augmentation to graphs is relatively under-explored. The complexity of non-Euclidean structures in graph data presents significant challenges that limit direct analogies to traditional augmentation operations on other types of data, such as images~\cite{CMC,SimCLR}, text~\cite{Eda} or videos~\cite{Videomix,VideoSurvey}. In machine learning applications that use graph data, the data typically include both graph structure (or edge features) and node features. Existing data augmentation techniques for graphs include adding or removing edges or nodes~\cite{GCA,graph_data_augmentation,graph_data_augmentation2,graph_data_augmentation3}, masking or shuffling node features~\cite{Deep,GraphCL,DeepGraphInfomax}, and sampling subgraphs~\cite{LA_GNN}. However, traditional graph augmentation techniques typically involve dropping nodes or scrambling edges, which may violate the physical semantics of the decay process. It is crucial to devise appropriate data augmentations for particle collision experiments.

To further improve the performance of data augmentation in representation learning tasks, we introduce a supervised contrastive learning (SCL) method.
Inspired by the recent success of self-supervised methods applied to images, learning graph-structured data has been increasing rapidly, especially for data augmentation-based contrastive methods.
As a typical self-supervised learning (SSL) method, contrastive learning has attracted increasing research interest~\cite{SimCLR,MOCO}. 
Recent works have explored graph contrastive learning method for molecular machine learning on various molecular property benchmarks, with promising results~\cite{wang2022molecular,sanchez2023cloome,fang2023knowledge}. However, the exploration of contrastive learning in particle physics is still limited.
To construct similar pairs and maximize the consistency between them in graph contrastive learning, existing methods rely on generic graph augmentation techniques. 
However, making modifications to the structure of the graph may not be suitable for application to particle decay reconstruction, as the addition or deletion on edges or nodes changes the nature and characteristics of the decay process.
Another issue is that self-supervised learning does not have explicit real labels for the input data during the learning process.
In contrast to self-supervised contrastive learning, which utilizes data augmentation to construct positive sample pairs and treats all other samples as negative samples, supervised contrastive learning introduces real label information about data and constructs correct positive and negative sample pairs while training.

\begin{figure*}[t]
    \centering
    \includegraphics[width=\linewidth]{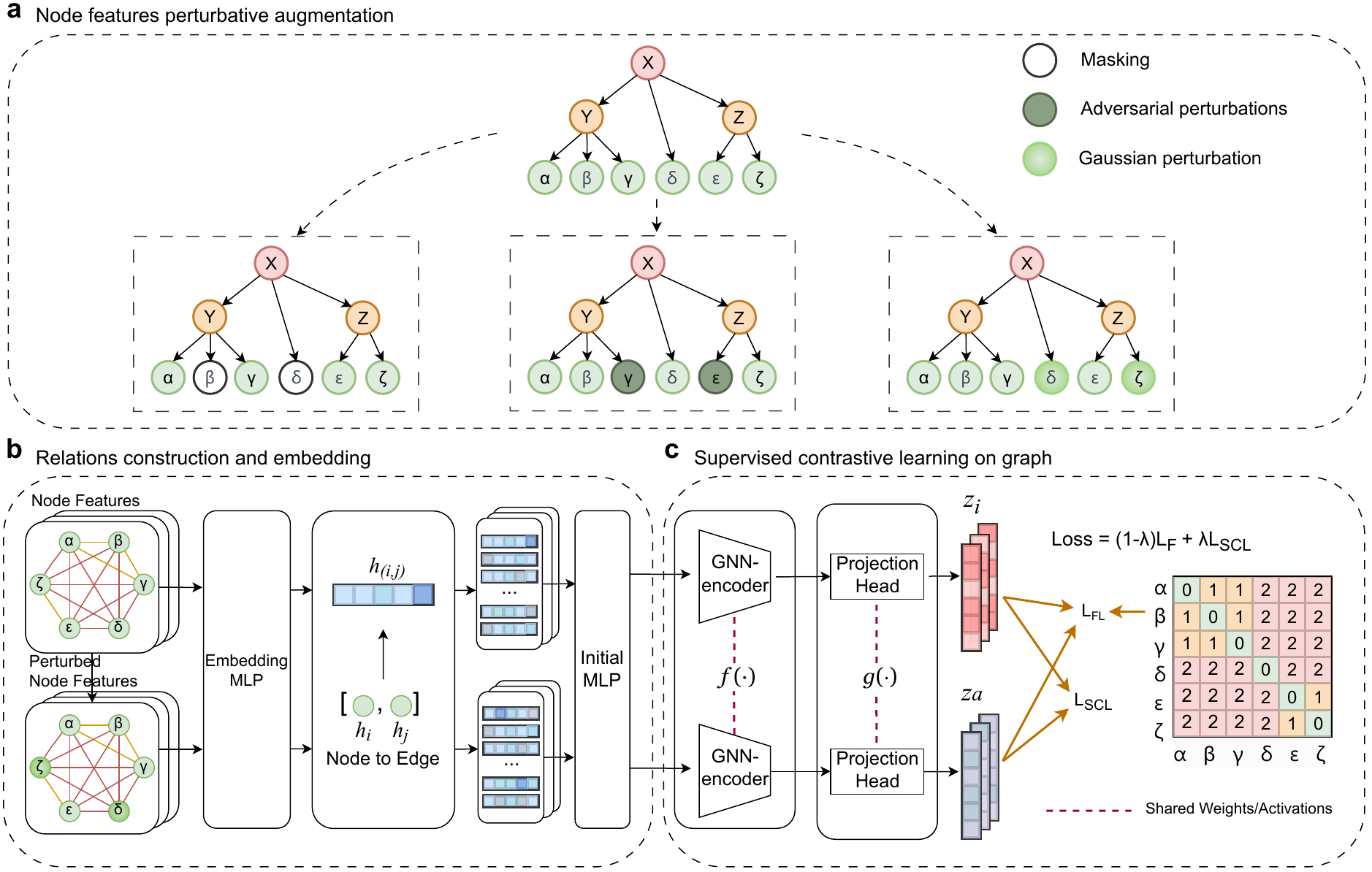}
    \caption{$\textbf{Overview of \Ours{}}$. 
    $\textbf{a}$: Node features perturbative augmentation. We use three perturbative augmentation methods: masking, adversarial perturbation and Gaussian perturbation to construct the augmented particle graph $\widetilde{\mathcal{G}}$. $\textbf{b}$: Relations construction and embedding. Extract the representations of $h$ and $\widetilde{h}$ between nodes in original particle graph $\mathcal{G}$ and the augmented particle graph $\widetilde{\mathcal{G}}$, using MLP-based message passing and aggregation. 
    $\textbf{c}$: Supervised contrastive learning on graph. 
    We establish the connection between different decay processes by constructing positive and negative sample pairs among them. The graph encoder and projection head is then trained to maximize the consistency between the same relations.}
    \label{overview}
\end{figure*}

To this end, we propose a novel Supervised graph Contrastive Learning with Perturbative Augmentation (\Ours{}), which utilizes a graph neural network to extract semantic features from the momentum and energy of particles \sw{for the}
particle decay reconstruction task.
An overview of \Ours{} and our contrastive learning scheme is shown in \fref{overview}.
We thoroughly evaluate the performance of \Ours{} on the Phasespace particle decay reconstruction dataset, demonstrating that it outperforms competing baselines \mymtx{without additional parameters}. We also conduct extensive \sw{ablation} experiments to validate the necessity of each \et{of the} components and to investigate the robustness and interpretability of \Ours{}.



\section{Methods}\label{sec:methods}
\mymtx{In the following sections, we will provide a detailed description of our methods, namely, the LCAG matrix formalism and the two new components that we propose: (1) node feature-based perturbative augmentation and (2) supervised contrastive learning on graph.}

\subsection{LCAG matrix formalism}

Arranged in chronological order, a series of particle decays can be naturally represented as a graph or, more specifically, as a tree with the initial particle being the root. The tree's leaves are composed of particles that reach the detector, while the intermediate nodes are the particles' ancestors, and the edges of the graph represent the parent-child relationship between particles. However, the complexity of directly predicting tree structures makes it infeasible in situations involving a large number of tree topologies. \mymtx{Since it is a common scenario in which the depth of the tree or the degrees of nodes within the tree are unknown in advance.} The Lowest Common Ancestor Generations (LCAG) matrix introduced in~\cite{LCAG} provides a way to represent tree structures that reduce the complexity to a manageable level, \mymtx{requiring only graph of size the number of leaf nodes of the tree.} In the following, \mymtx{we adopt notations of~\cite{LCAG}, where }a graph $\mathcal{G(V,E)}$ is a set of \sw{nodes} $\mathcal{V}$ with \sw{edges} $\mathcal{E}$ with relations (i.e. edge connections) defined by an adjacency matrix $\mathbf{A}\in\{0,1\}^{N\times N}$. 

\subsection{Node feature-based perturbative augmentations}\label{FeatPerturb}

\mymtx{To better utilize the particles' features to reconstruct the entire physical process, we propose to use node feature-based augmentation techniques.}
\et{Inspired by experimental effects such as material interactions and detector resolution, we introduce augmentations to our data by randomly perturbing the input features. This method has not been thoroughly investigated in the particle reconstruction context, and is intended to aid model generalizability and avoid over-dependence on the processes specific to the training dataset.}

\Fref{overview}a shows a snapshot of perturbative augmentation.
We use three graph data augmentation techniques: adversarial perturbations~\cite{FLAG}, Gaussian perturbations, and masking. Adversarial perturbation iteratively augments node features with gradient-based adversarial perturbations during training, thereby helping models generalize to out-of-distribution samples and boosting model performance at test time. Gaussian perturbation addresses the statistical variation of particle properties by adding Gaussian distribution perturbations to the features of a certain percentage of nodes to simulate the inherent experimental fluctuations. In particular, the masking technique randomly masks out the features of a certain percentage of particles to simulate the situation when certain measurements are not available. 
By applying these methods to \sw{a} graph, we \sw{can} obtain two views of \sw{this graph, namely} the original view and the perturbed view. \sw{In the following, we will introduce those three augmentation techniques.}

\paragraph{Node features adversarial perturbation}Adversarial data augmentation is an effective method to improve the robustness of neural networks. The adversarial data augmentation method on graph node features can be \sw{achieved} by the following steps:

First, we need to generate adversarial perturbations. Adversarial attack algorithms \mymtx{PGD~\cite{PGD}} can be used to generate adversarial perturbations, which are then added to the training data. This process is often formulated as the following min-max problem:
\begin{equation}
    \min\limits_\theta\mathbb{E}_{(\{\mathbf{X},\mathbf{Y})\sim\mathcal{D}}
    \left[\max\limits_{{\delta}\in\mathcal{P}}\mathcal{L_\mathit{FL}} \left (f_\theta \left ( \mathbf{X}+\delta  \right ),\mathbf{Y} \right )\right],\label{eq1}
\end{equation}
where $\mathcal{P}$ is the perturbation budget, \mymtx{randomly initialized by Gaussian perturbations on the node features}. $\mathbf{X}$ is the nodes feature matrix, $\mathbf{Y}$ is the label set of relations between nodes, and $\mathcal{L}$ is the objective function that we use focal loss for efficiency.

Second, the typical approximation of the inner maximization in \eref{eq1} is as follows:

\begin{equation}
    \delta_{t+1}=\delta_{t}+\mu\cdot \mathrm{sign}\left(\bigtriangledown_{\delta}\mathcal{L}\left(f_{\theta}\left(\mathbf{X}+\delta_{t}\right),\mathbf{Y}\right)\right),\label{eq2}
\end{equation}
where the perturbation $\delta$ is updated iteratively with the time step $t$. 

Third, to achieve \sw{higher} robustness, the iterative updating procedure is typically repeated $\mathit{m}$ ascent steps to generate the most detrimental perturbation.
Subsequently, the most severe noise $\delta_\mathit{m}$ is applied to the input feature, on which the model weights are optimized. \sw{Afterwards, }these perturbations \sw{will} be added to the original node features. The resulting transformation is: $\widetilde{\mathbf{X}}=\mathbf{X}+\delta_\mathit{m}$.

\paragraph{Node features Gaussian perturbation}To simulate the measurement errors that occur when detecting high-energy particles in experimental hardware, we propose to introduce perturbations to a certain percentage (e.g. 10\%) of nodes in the input graph during training. Specifically, we apply a certain level of perturbation $\delta$ to the node features, while leaving the graph structure unchanged. The perturbation $\delta$ is assumed to follow a Gaussian distribution with mean and variance equal to the mean $\mu$ and variance $\sigma^2$ of $\mathbf{X}$ and multiplied by a factor $\rho$ (e.g. 0.1). Formally, $\delta \sim \mathcal{N}(\mu, \sigma^2)$. The resulting transformation is: $\widetilde{\mathbf{X}}=\mathbf{X}+\rho\cdot\delta$.

\paragraph{Node features masking}On the node features masking, similar to Masked Autoencoder (MAE)~\cite{MAE} in image patch masking, we mask the feature matrix by randomly setting some elements of $\mathbf{X}$ to 0 during training. 

Formally, we first randomly sample a binary mask $p_i\sim{Bernoulli\left(1-\zeta \right)}$ for each node $n_i$. Second, we obtain the masked feature matrix $\widetilde{\mathbf{X}}$ by \sw{applying} each node's feature vector with its corresponding mask, i.e., $\widetilde{\mathbf{X}}_i=p_i\cdot{\mathbf{X}}_i$ where $\mathbf{X}_i$ denotes the $i^{th}$ row vector of $\mathbf{X}$.

\subsection{Supervised graph contrastive learning framework}
To further learn the representation of particle graphs, we employ supervised contrastive learning \sw{to better discriminate the original and augmented graphs.}
\sw{In particular, our} approach considers both the original and perturbed views of multiple particle decay processes, enabling the model to learn a more robust graph representation and improve its generalization and robustness.

After obtaining the \sw{augmented} graph and the embedding of its edges, we aim to incorporate them into training to enhance the model's understanding of the underlying \sw{graph characteristics}.
As shown in \fref{overview}c, to address this problem and to establish more meaningful connections across multiple decay processes, we employ supervised contrastive learning to train graph neural networks by maximizing the consistency between the original and augmented graphs. Given a small batch consisting of graphs corresponding to the decay processes of N particles, we create a set of 2N graphs and obtain the embeddings of their edges by converting their original particle graphs $\{\mathcal{G}_i\}_{i=1}^N$ to augmented particle graphs $\{\widetilde{\mathcal{G}}_i\}_{i=1}^N$ using node-feature-based perturbation of graph augmentation \sw{methods}.
Unlike the existing supervised contrastive learning method SupCon~\cite{SCL} where a batch contains only N images, a mini batch on the graph setting includes not only N graphs, but also M edges in each graph. \sw{Therefore,} \sw{within a mini-batch, }we consider edges with the same label as positive sample pairs and edges with different labels as negative sample pairs. We apply \sw{a} graph encoder $f(\cdot)$ to extract the edge embeddings $\{h_i\}_{i=1}^{NM}$and $\{\widetilde{h}_i\}_{i=1}^{NM}$ from the two graph views and apply the nonlinear projection network $g(\cdot)$ to map these embeddings into the space where the supervised contrastive loss is applied to obtain two new representations $\{z_i\}_{i=1}^{NM}$and $\{\widetilde{z}_i\}_{i=1}^{NM}$. Finally, the supervised contrastive loss is used to maximize the consistency between positive pairs while minimizing the consistency between negative pairs.
Note that unlike the existing supervised graph contrastive learning method ClusterSCL~\cite{ClusterSCL}, we construct sample pairs between edges instead of points, and ClusterSCL does not perform extra data augmentation.

\paragraph{Graph encoder and projection network}
Given the graph structure and node features, our goal is to learn a graph encoder $f(\cdot)$ that can obtain the representation of edges from an input graph. 
In our case, we utilize the encoder component of Neural relational inference (NRI)~\cite{NRI} as the graph encoder following~\cite{LCAG}, since our focus is to learn the hidden representation of relations. 

After obtaining representations of the edges using the graph encoder $f(\cdot)$, a non-linear transformation $g(\cdot)$ called the projection network is utilized to map both the original and augmented graph representations to a latent space where the supervised contrastive loss is calculated, as proposed in SupCon~\cite{SCL}. In this work, we employ a two-layer perceptron (MLP) with ELU activation function and batch normalisation to perform the projection network $g(\cdot)$.
Then, we propose a supervised graph contrastive learning objective, i.e., its loss function is a combination of focal loss and supervised contrastive loss on the labeled edges. Given $N$ augmented graphs under perturbative augmentation, we can naturally design a supervised graph contrastive loss for PASCL's graph based supervised learning.

\paragraph{Focal loss}
\sw{We use focal loss because we are targeting a multi-target classification task, where the numbers of different categories are extremely imbalanced. }
Specifically, in our task, it is more important to obtain correctness in the prediction of all LCAG entries than to obtain correctness in only a high-confidence subset as cross-entropy loss (CE) does. Therefore, we use focal loss (FL)~\cite{FocalLoss} instead of cross-entropy loss. We extend the focal loss to the case of graphs:
\begin{equation}
\mathcal{L_\mathit{FL}}=-\frac{ 1 }{2NM}\sum_{i=1}^{2N}\sum_{t=1}^{M}\alpha_t\left(1-p_t\right)^ \gamma\log\left(p_t\right),\label{eq5}
\end{equation}
where $\alpha_t$ is an adjustable scaling factor to balance the contribution of each class, $\gamma$ is an adjustable exponent parameter to down-weight the loss for well-classified examples, and $M$ denotes the number of edges in a graph. With $p_t$ representing the predicted probability of edge by the model, it can be rewritten as $CE(p_t)=-\log(p_t)$. The individual losses are averaged to produce the total loss. 

\paragraph{Supervised Contrastive Loss}
In the contrastive learning setting, we need to construct pairs of positive and negative samples. 
We propose a \sw{modified} supervised contrastive loss~\cite{SCL} that makes it applicable to \sw{our} graph data. First, the final representations $z$ of all the edges in the two views constitute positive pairs $(z_i,z_p)$, where $y_i=y_p$, while $z_i,z_a$ form negative pairs, where $y_i\neq y_a$, either in the same view and the other view. Secondly, different $N$ decay processes in the same batch are also used simultaneously to construct positive and negative sample pairs. It can process pairs of positive and negative samples in multiple decay processes simultaneously, and train the graph neural network to maximize the agreement between positive pairs and the discrepancy between negative pairs. Formally:

\begin{equation}
\begin{split}
    \mathcal{L_\mathit{SCL}}
    &=\sum_{i\in 2NM}\frac{-1}{\left|P\left(i\right)\right|}\sum_{p\in P\left(i\right)}\log_{}{\frac{\exp\left(z_i\cdot z_p/\tau\right)}{\sum_{a\in A\left(i\right)}^{}\exp\left(z_i\cdot z_a/\tau\right)}},\label{eq6}
\end{split}
\end{equation}
where the index $i$ is called the $anchor$, index $p$ is called the $positive$, and $A(i)\equiv 2NM\setminus \{i\}$ is the set of indices of all samples in two views distinct from $i$, while $P(i)\equiv \{p \in A(i):y_p=y_i\}$ is just for all positives, distinct from $i$. And $\tau$ is a scalar temperature parameter. 

\paragraph{Training and inference}\mymtx{Inspired by~\cite{SCL_fine-tuning}, we employ both the supervised classification loss in \eref{eq5} and the contrastive loss in \eref{eq6} during training. However, the differences are that we use focal loss instead of cross-entropy loss, and the supervised contrastive loss is used on graph-structured data instead of textual data.}
The final loss of \Ours{} is:
\begin{equation}
    \mathcal{L}=\left(\text{1}-\lambda\right)\mathcal{L}_\mathit{FL}+\lambda \mathcal{L}_\mathit{SCL},\label{eq7}
\end{equation}
where $\lambda$ is a scalar weighting hyper-parameter that controls the balance between the two losses. Specifically, when $\lambda=1$, it refers to supervised contrastive learning without considering focal loss. And when $\lambda=0$, it refers to supervised learning that only takes focal loss into account.

During the inference stage, we directly use the original node features for propagation and perform classification based on the final edge embeddings. This is justified because we scale the edge embeddings from the augmented views $\widetilde{\mathbf{E}}_i$ to guarantee its expectation to match those from the original view $\mathbf{E}_i$.

\section{Experimental setup}\label{sec:setup}

\subsection{Training and test data}
In our study, we use a generated Phasespace dataset~\cite{dataset}.
This dataset consists of simulated synthetic particle decays generated using the PhaseSpace library~\cite{phasespace_library}, which is designed to model kinematically-allowed particle decay trees that are governed by only the law of energy and momentum conservation.
\sw{Our used} dataset consists of 200 topologies, each trained with 4,000 or 16,000 samples, as appropriate to the specific experiment, and validated and tested with half the training sample size.


\subsection{Implementation details}

We train and evaluate all models using the aforementioned simulated dataset. During the training phase, we conduct 50 epochs with a batch size of 64, dropout rate of 0.3, random seed of 42, while also applying class weights. We utilize the Adam optimizer with a learning rate of 0.001, and set the feed-forward dimensions of the models to 512.
For the perturbative augmentations, the step size $\mu$ we set to the empirical value 0.1 and $\zeta\in\{0.1, 0.3, 0.5, 0.7, 0.9\}$.
For the SupCon Loss, we conduct a grid-based hyperparameter sweep for $\lambda\in\{0.1, 0.3, 0.5, 0.7, 0.9\}$ and $\tau\in\{0.07, 0.1, 0.4\}$. We find that the models with the highest test accuracy in all experimental setups overwhelmingly use the hyperparameter $\tau=0.07$. At the same time, the hyperparameter $\lambda$ \sw{is set to 0.1 since it }does not \sw{have large impact on} the test accuracy.
For the focal loss, we set the adjustable exponential parameter $\gamma$ to 2. To ensure a fair comparison, our backbone references the optimal settings of the baseline NRI~\cite{LCAG}, which consist of two NRI blocks and an initial/final MLP layer.
The NRI block employed sequences of MLPs containing two linear layers with ELU activations and batch normalisation.
\Ours{} is implemented using pytorch and runs on an Ubuntu server with an NVIDIA GeForce RTX 2080Ti graphics processor.

\subsection{Baselines}We present a comparison of \Ours{} with six supervised/self-supervised learning baselines for edge-level tasks, which include NRI~\cite{NRI}, FNRI~\cite{fNRI}, GCN~\cite{GCN}, GAT~\cite{GAT}, EdgeConv~\cite{EdgeConv} and GCA~\cite{GCA}, and two experimental models of ablation without supervised contrastive loss and without perturbative augmentation. For each supervised learning method, we use \textit{focal loss} for end-to-end training. In particular, for the node-based methods, instead of associating two nodes before the neural network, we concatenate two nodes after the neural network as edge features and apply an MLP to obtain the class probability. GCA is a self-supervised graph contrastive learning method, we modify GCA's node-based loss function to edge-based to align it with our task, and use our three proposed methods for data augmentation.


\subsection{Evaluation Metrics}Since this is a multi-target classification task, i.e., each LCAG entry has a target, it is more important to obtain the correctness of the predictions for all LCAG entries than to obtain the correctness of the predictions for only a subset with high confidence. Therefore, following~\cite{LCAG}, the performance is reported on the perfectLCAG metric to assess the correctness of LCAG predictions, i.e., the ratio of correctly predicted LCAGs to the total number of samples.


\section{Results}\label{sec:results}


\begin{table*}[t]
  \centering
  \small
  \caption{Test performance of different methods, each model is trained on the Phasespace dataset with 16,000 samples per topology. The first 5 models are supervised learning methods, and the middle 2 are contrastive learning methods, while the last 2 are ablation studies. Underlined numbers indicate the best score at each complexity.}\label{table1}
  \begin{tabularx}{\textwidth}{c*{1}{>{\raggedright\arraybackslash}}c*{14}{>{\centering\arraybackslash}X}}
    \toprule
    {} & \multicolumn{15}{c}{Average perfectLCAG (\%) score for leaves up to and including} \\
    \cmidrule{2-16}
    Model & 2 & 3 & 4 & 5 & 6 & 7 & 8 & 9 & 10 & 11 & 12 & 13 & 14 & 15 & 16 \\
    \midrule
    

    NRI & \underline{100.0} & 99.0 & 98.7 & 97.5 & 95.9 & 90.9 & 86.3 & 76.9 & 67.8 & 62.4 & 58.1 & 56.3 & 55.5 & 55.0 & 53.9 \\
    FNRI & \underline{100.0} & 98.6 & 98.3 & 97.9 & 96.1 & 92.9 & 90.9 & 85.7 & 78.6 & 74.6 & 70.3 & 68.4 & 67.6 & 67.0 & 65.6 \\
    GCN & \underline{100.0} & 65.6 & 47.2 & 33.4 & 27.0 & 18.7 & 14.3 & 10.7 & 8.5 & 7.5 & 6.9 & 6.6 & 6.5 & 6.4 & 6.3 \\
    GAT & \underline{100.0} & 63.7 & 45.8 & 31.6 & 25.6 & 17.8 & 13.6 & 10.2 & 8.1 & 7.1 & 6.5 & 6.3 & 6.2 & 6.1 & 6.0 \\
    EdgeConv & \underline{100.0} & 97.8 & 94.9 & 90.6 & 82.2 & 66.5 & 54.3 & 40.9 & 32.9 & 29.0 & 26.6 & 25.8 & 25.4 & 25.2 & 24.7 \\
    \midrule 
    GCA & \underline{100.0}&98.6&98.5&98.3&96.5&93.0&90.5&84.4&76.5&71.6&67.2&65.3&64.4&63.9&62.6\\
    \midrule
    \Ours{}&\underline{100.0} & \underline{99.1} & \underline{98.9} & \underline{98.7} & \underline{97.3} & \underline{94.3} & \underline{92.3} & \underline{87.2} & \underline{80.5} & \underline{76.2} & \underline{71.9} & \underline{70.0} & \underline{69.2} & \underline{68.6} & \underline{67.3}\\
    \midrule
    w/o SCL& \underline{100.0} & 98.7 & 98.7 & 98.6 & 97.2 & 94.2 & 92.0 & 86.7 & 77.6 & 72.5 & 67.9 & 65.9 & 65.1 & 64.5 & 63.2\\
    w/o Aug &  \underline{100.0} & 99.0 & 98.8 & 98.6 & 96.9 & 94.0 & 91.4 & 85.8 & 78.2 & 73.9 & 69.9 & 68.0 & 67.1 & 66.5 & 65.2\\
    \bottomrule
  \end{tabularx}
\end{table*}

\subsection{Results and Analyses}\Tref{table1} reports the overall accuracy (average perfectLCAG score, \%) of the physical particle decay reconstruction task. Our experimental results show that our proposed supervised graph contrastive learning model PASCL achieves state-of-the-art improvement in accuracy.

We make the following observations. The results demonstrate that PASCL outperforms all supervised/self-supervised baseline models at all number of leaves on the Phasespace dataset. Particularly, compared to the NRI decaying to 86.3\%, PASCL can still achieve an accuracy of 92.3\% for trees with no more than 9 leaves. For trees with no more than 16 leaves, average perfectLCAG score reaches 67.3\%, which is a significant improvement \sw{(13.4\%)} over NRI.

Similar improvements can also be observed on contrastive learning. PASCL also achieves significant improvement \sw{over} the baseline GCA, \sw{which also uses} contrastive learning. Average perfectLCAG score of PASCL exceeds GCA on all number of leaves, and still has 4.7\% improvement when the number of leaves is up to 16. The improvements on accuracy indicate that our proposed supervised contrastive learning loss has stronger learning ability on the task of learning tree structure from leaves.

These results indicate that PASCL is able to better capture the information in the graph structure when learning node and edge representations, which can improve the performance of particle decay reconstruction.

\subsection{Investigation of perturbative augmentation} 

\begin{table*}[t]
  \centering
  \caption{Test performance of different perturbative augmentation methods on \Ours{}, train on the Phasespace dataset with 4,000 samples per topology. Underlined numbers indicate the best score at each complexity.}\label{table2}
  \begin{tabularx}{\textwidth}{c*{1}{>{\raggedright\arraybackslash}}c*{14}{>{\centering\arraybackslash}X}}
    \toprule
    {} & \multicolumn{15}{c}{Average perfectLCAG (\%) score for leaves up to and including} \\
    \cmidrule{2-16}
    {Model} &  2 & 3 & 4 & 5 & 6 & 7 & 8 & 9 & 10 & 11 & 12 & 13 & 14 & 15 & 16 \\
    \midrule
    NRI & \underline{100.0} & 98.9 & 97.7 & 97.5 & 94.7 & 89.9 & 83.6 & 72.8 & 63.3 & 57.7 & 53.4 & 51.8 & 51.0 & 50.5 & 49.5 \\
    adversarial &  \underline{100.0} & \underline{99.0}&\underline{98.7}&\underline{98.6}&\underline{96.5}&\underline{92.7}&\underline{89.8}&\underline{82.6}&\underline{74.6}&\underline{70.1}&\underline{65.7}&\underline{63.8}&\underline{62.9}&\underline{62.4}&\underline{61.1}\\
    Gaussian& \underline{100.0}& 98.8&98.4&97.7&95.2&91.2&88.3&81.1&73.0&68.5&64.2&62.3&61.4&60.9&59.7\\
    mask& \underline{100.0}&98.9&98.1&97.6&95.3&91.8&88.8&80.6&72.0&66.7&62.3&60.4&59.6&59.1&57.9\\
    \bottomrule
  \end{tabularx}
\end{table*}

\begin{figure*}[t]
    \centering
    \includegraphics[width=1\linewidth]{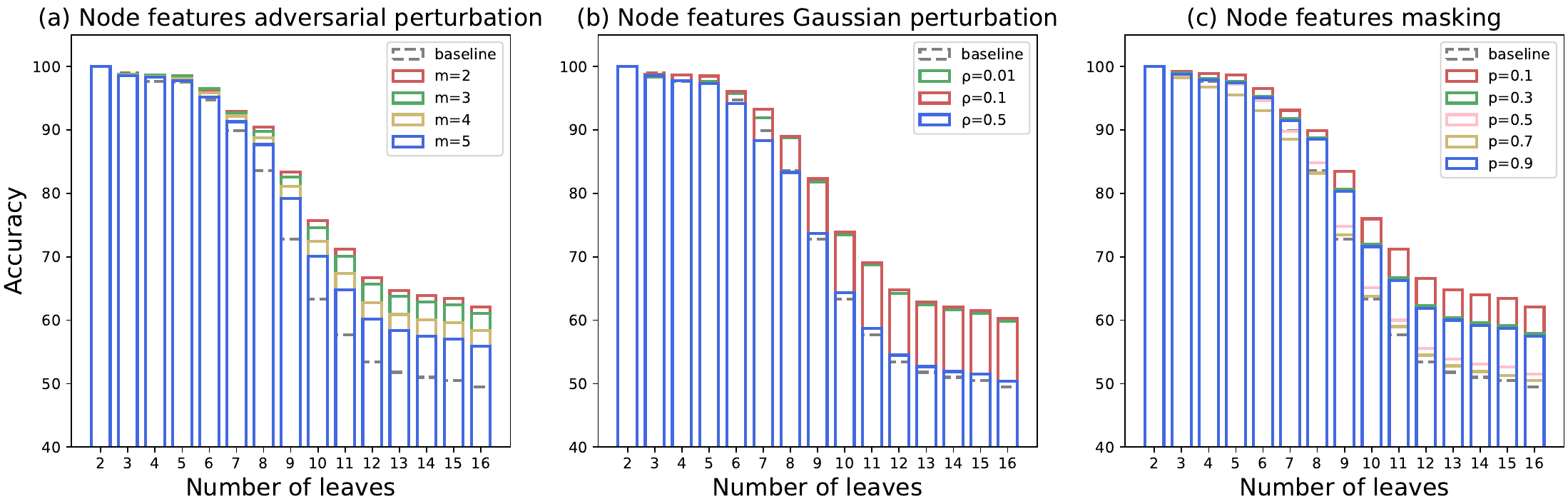}
    \caption{\textbf{Performance of \Ours{} with different perturbative augmentation methods.} We vary the degree of different perturbations in data augmentation and report the corresponding performance trends across different parameters, represented by different colors. The red represents the best performing model, blue represents the worst, and the gray dashed line represents the baseline. The horizontal axis represents the number of leaves up to and including, and the vertical axis represents the accuracy in the classification task (higher is better). Results are expressed as the average of ten separate runs.}
  \label{ablation}
\end{figure*}

Perturbative augmentation is crucial in \Ours{} framework because it provides rich and diverse inputs to the model, thus helping it to better capture the complexity and abstract features of the data. By applying small perturbations and transformations to the raw data, perturbative augmentation extends the dataset on which the model is trained, giving it greater generalization capabilities. This is particularly important for contrastive learning, which relies on distinguishing between positive and negative samples, and enrich data variations can improve the model's ability to perceive the variability between different samples.

We employ three perturbative data augmentation methods. For practical efficiency, our experiments are conducted on phasespace dataset consisting of 4,000 samples per topology for both training and validation, and 2,000 samples for testing. \Tref{table2} and \fref{ablation} present the different methods and the results of hyperparameter search for each method.

Our experimental results demonstrate that each of the methods exhibits varying degrees of improvement compared to the NRI baseline. \Tref{table2} displays the optimal results achieved by each method. 
We find that \sw{with} the adversarial and Gaussian perturbations, our model performs best when the proportion of perturbed nodes is 10\% and the scaling factor $\rho$ is 0.1, which is consistent with observations in the masking.
Notably, \sw{\Ours{} with} adversarial perturbation method outperforms NRI by 11.6\% when ascent steps $m=2$, while the other two methods show improvements of 10.2\% and 8.4\%, respectively.

\Fref{ablation} illustrates the results of the hyperparameter search, with the gray dashed line representing the baseline, solid lines representing our models, and the red solid line denoting the best-performing configuration. For detailed parameter settings, please refer to the \sref{FeatPerturb}.

\subsection{Investigation of PASCL representation}
\begin{figure*}[t]
    \centering
    \includegraphics[width=1\linewidth]{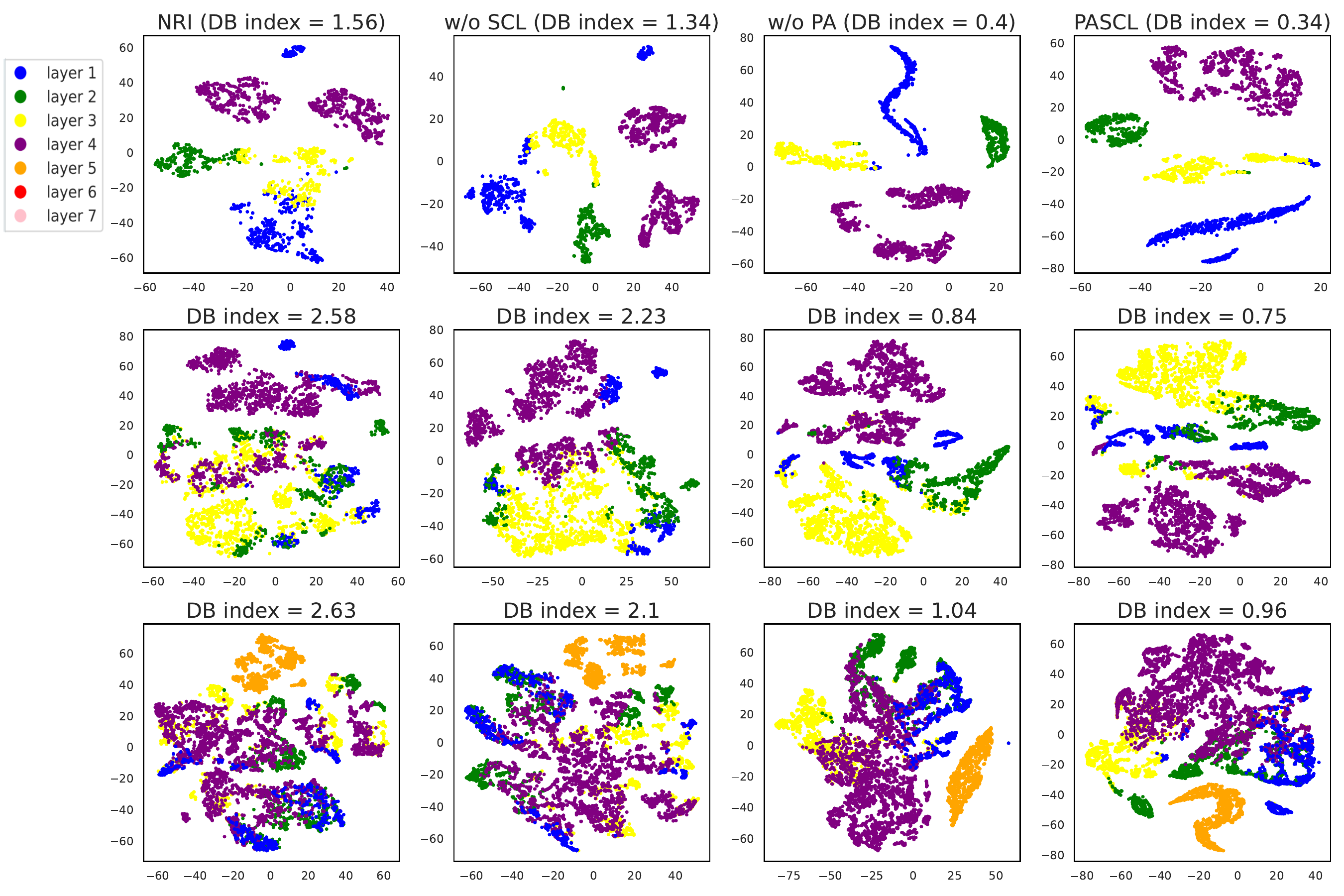}
    \caption{\textbf{Alignment analysis}. We show t-SNE visualizations of inter-particle relationship representations during 50 random decay processes to investigate the correlation of splitting at the same level. The colors represent the level of splitting (i.e. the value of the edge in the truth LCAG matrix). From top to bottom are the cases of 8, 12, and 16 leaf nodes, respectively.}
    \label{t-sne}
\end{figure*}
We also assess the validity of the model by investigating the representation space they \sw{have learned}. The quality of the representation space can be assessed by alignment, where embeddings of similar samples should be mapped \sw{into closer space}. In \fref{t-sne}, we compare the final embedding of relations derived by our method with those derived by a supervised modeling approach NRI~\cite{LCAG}.

We perform visual representations of the final embeddings of different relations by t-distributed stochastic neighbour embedding (t-SNE)~\cite{t-SNE} to test whether the same relations have similar representations.
Relationships represent the layers in which two particles split and are important information in the particle decay process. We select three different decay processes ([8, 12, 16] leaf nodes) from the phasespace dataset and use different colors to distinguish different relations between particles.
As shown in \fref{t-sne}, the model with only supervised loss is unable to distinguish between these inter-particle relations, while the perturbative augmentation and supervised contrastive methods provide a slight improvement.
In contrast, the more discriminative clusters generated by \Ours{} have the lowest Davies–Bouldin (DB) index~\cite{DB_index}, which means a better clustering effect in the latent space.

\subsection{Ablation study}

We conduct an ablation study to examine the contributions of different components in \Ours{}.

\begin{itemize}
    \item \textbf{Without supervised contrastive learning (SCL)}: We only use the supervised classification loss, i.e., $\lambda=0$.
    \item \textbf{Without perturbative augmentation (PA)}: We construct positive and negative sample pairs only by labels.
\end{itemize}

\Tref{table1} summarizes the results of the ablation study, from which we can have two observations. First, the performance of all \Ours{} variants with some components removed is significantly degraded compared to the full model, indicating that each component of the design contributes to the success of \Ours{}. Second, the crippled \Ours{} outperforms the five supervised/self-supervised baselines, except for FNRI, which proves the importance of the proposed supervised graph contrastive learning technique for relational inference.


\section{Conclusion}\label{sec:conclusion}
In this work, we propose \Ours{}, a novel approach that improves the prediction of inter-particle relations through supervised graph contrast learning with perturbative augmentation. Experiments show that \Ours{} achieves superior reconstruction accuracy and learns more discriminative feature representations on the particle decay simulation phasespace dataset with an equivalent number of parameters. Our work is dedicated to better understanding and analyzing uncertainties in high-energy particle decay processes through perturbation data augmentation, while supervised graph contrast learning allows for more meaningful connections between multiple different decay processes.

Although \Ours{} has shown promising success in improving the accuracy of particle decay reconstruction, there are still some limitations. The applicability of our method in the field of high-energy particle physics requires broader validation and application, including different types of particle decay processes and more complex experimental scenarios.
To this end, we propose some directions for future work. First, we will continue to improve the model's ability to learn relational representations to improve the model's performance when there are more leaf nodes. Second, continue to extend the applicability of \Ours{} by validating it against more particle decay experimental datasets to ensure its generalizability across different experiments.
Finally, investigating the interpretability of the \Ours{} model and the learned representations can provide meaningful insights into the study of particle properties and event reconstruction.

\section*{Data availability}
The particle decay processes on a simulated physics dataset used in this work are available in the URL/DOI:~\url{https://doi.org/10.5281/zenodo.6983258}~\cite{dataset}.

\section*{Code availability}The source code is available at~\url{https://github.com/lukSYSU/PASCL}.

\section*{Acknowledgements}
This work is supported by the MBZUAI-WIS Joint Project.


\clearpage


\newcommand{\newblock}{}
\bibliographystyle{unsrt}
\bibliography{bibliography}

\end{document}